# Repeating Patterns in Linear Programs that express NP-Complete Problems

Deepak Ponvel Chermakani, IEEE Member
deepakc@pmail.ntu.edu.sg d.chermakani@sms.ed.ac.uk deepakc@usa.com deepakc@myfastmail.com

***Abstract: -*** **One of my recent papers transforms an NP-Complete problem into the question of whether or not a feasible real solution exists to some Linear Program. The unique feature of this Linear Program is that though there is no explicit bound on the minimum required number of linear inequalities, which is most probably exponential to the size of the NP-Complete problem, the Linear Program can still be described efficiently. The reason for this efficient description is that coefficients keep repeating in some pattern, even as the number of inequalities is conveniently assumed to tend to Infinity. I discuss why this convenient assumption does not change the feasibility result of the Linear Program. I conclude with two Conjectures, which might help to make an efficient decision on the feasibility of this Linear Program.**

## 1. Introduction

### 1.1 Previous related development on the univariate Polynomial

Given a univariate integer Polynomial Q(x) of degree $d$, it has been proved [1] that Q(x) does not have a positive real root, if and only if, Q(x) can be multiplied by some Polynomial P(x) with positive coefficients, so as to produce a resultant Polynomial with positive coefficients. Here we assume that both the coefficients of *1* and of $x^d$ in Q(x) are positive, and if not, then we assume without loss of generality that Q(x) is simply replaced with $Q^2(x)$. The minimum required degree of P(x), has been shown to be bounded by some function of the smallest non-zero imaginary roots of Q(x) [1]. If we denote $P(x) = p_0 + p_1 x + p_2 x^2 + ... + p_N x^N$, we can obtain a set of Linear Inequalities by stating that there exist non-negative real coefficients $<p_0, p_1, ... p_N>$, such that every coefficient of P(x)Q(x) is non-negative, and that the coefficient of 1 in P(x)Q(x) is greater than or equal to 1. This change from *strictly positive* inequalities to inequalities of *non-negativity* is achieved by dividing throughout by the product of the coefficient of 1 in Q(x) with the coefficient of 1 in P(x) that was originally assumed to be positive. Thus, the Problem of deciding whether or not Q(x) has a positive real root, can be expressed as a Standard Linear Programming (LP) Feasibility Problem. A feasible LP Solution exists, if and only if, Q(x) does not have a positive real root [2]. For example, consider $Q(x) = (x-1)^2 (x-2)^2 + 1 = x^4 – 6x^3 + 13x^2 – 12x + 5$. Our corresponding LP is defined by the following inequalities, in addition to the non-negativity constraints on all variables, i.e. $p_i \geq 0$, for all $i$ as integers in $[0,N]$:

$$
\begin{array}{rcl}
5p_0 & \geq & 1 \\
-12p_0 + 5p_1 & \geq & 0 \\
13p_0 - 12p_1 + 5p_2 & \geq & 0 \\
-6p_0 + 13p_1 - 12p_2 + 5p_3 & \geq & 0 \\
1p_0 - 6p_1 + 13p_2 - 12p_3 + 5p_4 & \geq & 0 \\
1p_1 - 6p_2 + 13p_3 - 12p_4 + 5p_5 & \geq & 0 \\
1p_2 - 6p_3 + 13p_4 - 12p_5 + 5p_6 & \geq & 0 \\
1p_3 - 6p_4 + 13p_5 - 12p_6 + 5p_7 & \geq & 0 \\
1p_4 - 6p_5 + 13p_6 - 12p_7 + 5p_8 & \geq & 0 \\
\cdots\cdots\cdots\cdots\cdots & & \\
\cdots\cdots\cdots\cdots\cdots & & \\
1p_{N-7} - 6p_{N-6} + 13p_{N-5} - 12p_{N-4} + 5p_{N-3} & \geq & 0 \\
1p_{N-6} - 6p_{N-5} + 13p_{N-4} - 12p_{N-3} + 5p_{N-2} & \geq & 0 \\
1p_{N-5} - 6p_{N-4} + 13p_{N-3} - 12p_{N-2} + 5p_{N-1} & \geq & 0 \\
1p_{N-4} - 6p_{N-3} + 13p_{N-2} - 12p_{N-1} + 5p_N & \geq & 0 \\
1p_{N-3} - 6p_{N-2} + 13p_{N-1} - 12p_N & \geq & 0 \\
1p_{N-2} - 6p_{N-1} + 13p_N & \geq & 0 \\
1p_{N-1} - 6p_N & \geq & 0
\end{array}
$$

In the above, only the top *4* and the bottom *4* inequalities have a unique structure, while the remaining inequalities have a repeating structure. Here, *4* happens to be the degree of Q(x), but in general, if $N$ and $d$ represent the degrees of P(x) and Q(x) respectively, we will have $2d$ inequalities with a unique structure, and $N-2d$ inequalities with a repeating structure. This makes it possible to give an efficient definition to the inequalities, even as $N$ is conveniently assumed to tend to infinity. Setting $N$ to tend to a number larger than required, does not cause any harm because, if Q(x) does not have a positive real root and if $r$ denotes the minimum required degree of P(x) such that P(x)Q(x) has non-negative coefficients, then an LP Solver can choose $p_i = 0$ for all $i$ as integers in $[r+1, N]$. Also, if Q(x) has a positive real root, then by Descartes Rule of Signs, there cannot exist a real Polynomial P(x), such that P(x)Q(x) has non-negative coefficients. The figure below plots feasibility ('1' means that a feasible LP solution exists, while '0' means that there is no feasible LP solution) versus the degree of P(x).

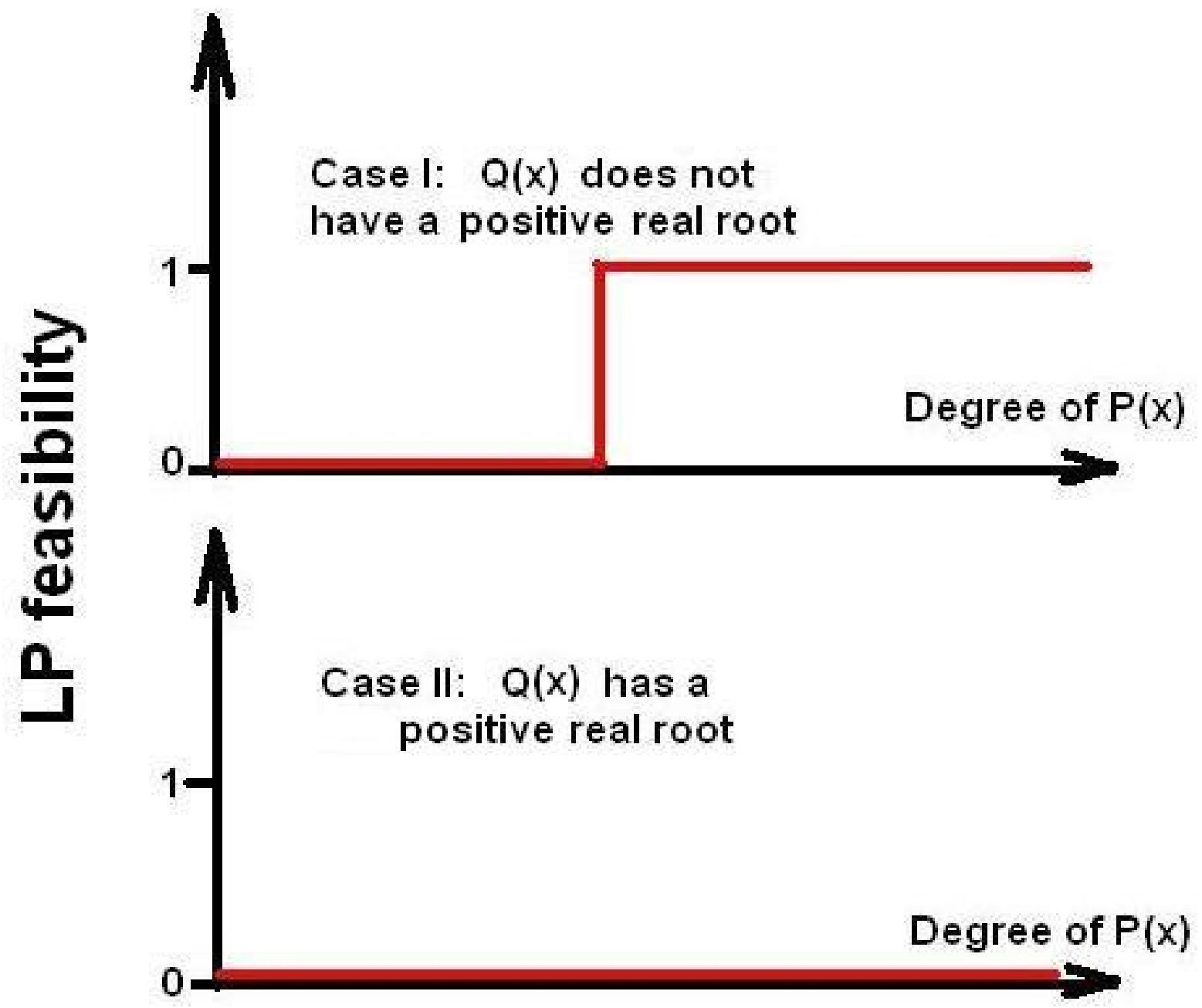

### 1.2 Expressing 3-SAT as a Linear Program, via the mentioned development on the univariate Polynomial

3-SAT is a well-known NP-Complete Problem that aims to decide on whether or not a Boolean expression is satisfiable, with a maximum of 3 literals per clause. Let us denote a 3-SAT instance to have $k$ clauses and $u$ binary variables.

3-SAT has been transformed into the problem of deciding real root existence for a univariate Polynomial Q(x) [4]. So going by the logic mentioned in the Section 1.1, we can obtain the corresponding LP expressing the 3-SAT instance. However, it is difficult to study this LP because of the following three properties of Q(x) obtained from the 3-SAT instance:

1) the degree of Q(x) is exponential to $k$ and $u$
2) Q(x) is usually not sparse, though it can be efficiently expressed as a Straight Line Program (A Straight Line Program is a set of instructions to describe a Polynomial)
3) the coefficients of Q(x) are exponential to $k$ and $u$

### 1.3 Previous related development on the multivariate Polynomial

Given a $u$-variate ($u > 1$) integer Polynomial Q, it has been shown that not all instances of Q (without a real root) can be multiplied by another Polynomial, to give a resultant Polynomial with positive coefficients [3]. The same author also described a transformation that maps every instance of Q without a real root, in the standard Simplex and the standard Hypercube, onto some Polynomial with positive coefficients [3]. However, it is to be noted that there also exist Polynomials with real roots that can be mapped onto Polynomials with positive coefficients, by this transformation.

## 2. The New Development on the multivariate Polynomial

To the best of my knowledge, Theorem-7 of my paper [2] is the first robust theoretical development for a $u$-variate ($u > 1$) integer Polynomial Q, where Q is constrained to either have integer roots in a bounded region, or not have a real root. It can be inferred that Theorem-7 can be generalized over a finite set of positive real numbers as follows.

### 2.1 Generalization of Theorem-7 of my paper [2] over a finite Set of positive Real numbers

Let N be a finite Set of positive Real numbers, i.e. $N = \{n_1, n_2, \ldots n_M\}$. Let a multivariate Polynomial Q defined in $u$-variables $x_1, x_2, \ldots x_i, \ldots x_u$, be such that exactly one of the following two statements, is true:

1) "Q = 0" implies that $x_i \in N$, for all integers i in [*1*,*u*]
2) Q does not have a real root

Next, define constant Polynomials $P_1, P_2, \ldots P_i, \ldots P_u$, where $P_i = ((x_i - n_1)(x_i - n_2)\ldots(x_i - n_M))^2$. Then the generalization of the Theorem is that Q does not have a real root, if and only if, there exist real *u*-variate Polynomials $K, K_1, K_2, \ldots K_u$, such that the following Polynomial has positive coefficients: $- QK + P_1K_1 + P_2K_2 + \ldots + P_uK_u$.

### 2.2 Expressing 3-SAT as a Linear Program, via the new development

The three step procedure in Section-4 of my paper [2] derives a *u*-variate Polynomial Q from a 3-SAT instance, such that this Q satisfies exactly one of the previous two statements mentioned in Section 2.1, where the finite Set N = {1,2}. As a result of Theorem-7, we can now transform 3-SAT into a Polynomial with positive coefficients, if and only if, it is not satisfiable.

We follow a similar logic mentioned in the Section 1.1, in which we transformed the problem of deciding real root existence for a univariate Polynomial into a feasibility problem in a LP. We define the real polynomials $K, K_1, K_2, \ldots K_u$ in terms of variable coefficients, just like how we defined P(x) in Section 1.1. But there is a small difference here, because the coefficients of P(x) were non-negative, but those of $K, K_1, K_2, \ldots K_u$ are real. For example, assuming that Q is bivariate, then define the Polynomial $K_2 = K_{2_00} + K_{2_01}x_2 + K_{2_10}x_1 + K_{2_02}x_2^2 + K_{2_11}x_1x_2 + K_{2_20}\, x_1^2 + \ldots + K_{2_ij}\, x_1^i x_2^j + \ldots + K_{2_AB}\, x_1^A x_2^B$, where A and B are natural numbers tending to infinity, and where $K_{2_ij}$ is a real variable, for each of the non-negative integers *i* and *j*, where *i* represents the power of $x_1$ and *j* represents the power of $x_2$. Note now that a real variable can easily be represented as two LP variables constrained to be non-negative, for example, replace $K_{2_ij}$ with $K_{2_ij_positive} - K_{2_ij_negative}$, where both variables $K_{2_ij_positive}$ and $K_{2_ij_negative}$ are non-negative. Similarly define the Polynomials K and $K_1$ with variable coefficients. Next, we state that every term of the resulting expression, is non-negative. Finally we state that the coefficient of 1 in the resulting expression is greater than or equal to 1.

Thus, Theorem-7 of my paper [2] can be viewed as transforming any instance of 3-SAT into a feasibility problem in a LP. A feasible LP solution exists, if and only if, the 3-SAT instance is not satisfiable. It can be inferred that this LP will have the following four properties:

1) the number of unique coefficients in each inequality is bounded by *27k*,
2) the magnitude of each these coefficients is bounded by *64k*,
3) the number of monomial terms in each of the inequalities is bounded by *ku*, and,
4) the minimum required number of linear inequalities (which is proportional to the required degrees of the Polynomials $K, K_1, K_2, \ldots K_u$) is most probably exponential to *k* and *u* in the worst case (though I proved that this number exists, I do not know how it grows). Next, even if the number of linear inequalities is conveniently assumed to tend to infinity (which will not change the feasibility result for the LP), the entire LP can still be defined efficiently, by the same logic described in Section 1.1.

As the best LP algorithms are weakly-polynomial, the above mentioned properties of this LP make it more attractive for determining feasibility, compared to the LP derived from the univariate approach mentioned in Section 1. This LP too has a repeating structure, which would take up too much space to show on this paper. However, one can verify this and conclude that there is such a repeating structure, which is more complex and beautiful than the simple repeating structure of the LP of Section 1.

## 3. Conjectures that might help in proving P=NP

Given the LP formulations obtained, either from the univariate Polynomial derived from 3-SAT as described in Section-1, or from the multivariate Polynomial derived from 3-SAT as described in Section-2, one could identify Patterns, and investigate the situations when we are able to make an efficient decision on feasibility for the LP.

Mentioned below are some of the techniques one could use to investigate how to make efficient decisions on the feasibility problem for the LP.

### 3.1 Extrapolation of the optimal Artificial Variables from the LP solutions

One way of doing this would be to study patterns in the solutions obtained by an LP Solver, as the degrees of the multiplying Polynomials (i.e. P(x) in the case of the approach in Section-1, or the Polynomials $K, K_1, K_2 \ldots K_u$ in the case of the approach in Section-2) are iteratively increased. Patterns can be studied in the behaviour of the *reduced cost*s of the optimal variables, and the optimal values of the *artificial variables* introduced into the LP. Note that the *artificial variable* technique in LP, is to introduce positive artificial variables into the inequalities, and then the LP is concluded to be feasible, if and only if, all artificial variables become zero, at the end of the LP runs. Below is a diagram of how I expect the optimal values of the artificial variables to behave, decreasing continuously with respect to the degrees of the multiplying Polynomials. So by getting the initial few values of the artificial variables, one could extrapolate and determine how future values of the artificial variables could behave.

### 3.2 Conjecture on the asymptotic convergence of the optimal Artificial variables

Another way would be to develop a novel approach to determine feasibility, as the degrees of the multiplying Polynomials are conveniently assumed to tend to infinity, by using concepts of how some sequences converge. Note that like the univariate case described in Section-1, it is harmless to use higher-than-required degrees of the multiplying Polynomials K, $K_1$, $K_2$...$K_u$ to obtain the Linear Program. Also, the Linear Program of Section 2 can be defined efficiently even though the degrees of K, $K_1$, $K_2$...$K_u$ are conveniently assumed to tend to infinity. Probably, knowledge of how some sequences converge asymptotically [5], would help in this. I conjecture that the summation of optimal artificial variables should:

1) Become equal to zero, if Q does not have a real root, as the degrees of the multiplying Polynomials tend to infinity
2) Asymptotically tend to a positive real, if Q = 0, as the degrees of the multiplying Polynomials tend to infinity

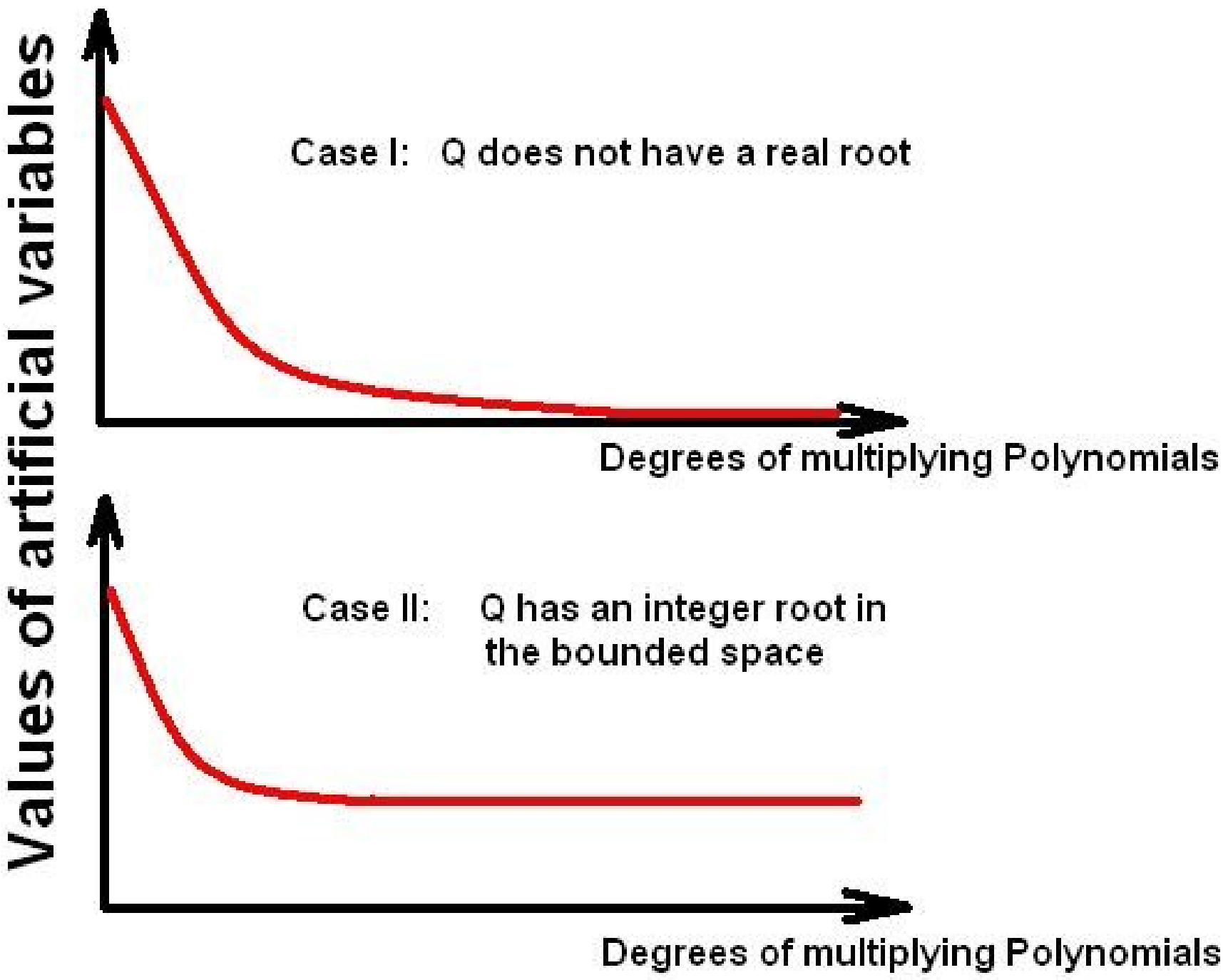

**References**

[1] D.R. Curtiss, *Recent Extensions of Descartes Rule of Signs*, Annals of Mathematics, Volume 19, No 4, pp 251-278, Jun-1918.
[2] Deepak Ponvel Chermakani, *Another approach to decide on real root existence for univariate Polynomials and a multivariate extension for 3-SAT*, arXiv:0803.0018v5, Sep-2008.
[3] David Handelman, *Positive Polynomials and product type actions of Compact Groups*, Memoirs of the American Mathematical Society, Number 320, Volume 54, Mar-1985.
[4] Daniel Perrucci, Juan Sabia, *Real root s of univariate Polynomials and Straight Line Programs*, Volume 5, Issue 3, Journal of Discrete Algorithms, pages 471-478, Sep-2007.
[5] David Moews, *Asymptotic behaviour of Rauzy's Sequence*, http://djm.cc/dmoews/rauzy.pdf, Aug-2002.

**About the Author**

I, Deepak Ponvel Chermakani, have written this paper, out of my own interest and initiative, during my spare time. I am currently a student at the University of Edinburgh UK (www.ed.ac.uk), where since Sep-2009, I have been enrolled in a fulltime one year Master Degree course in *Operations Research with Computational Optimization*. In Jul-2003, I completed a four year fulltime four year Bachelor Degree course in *Electrical and Electronic Engineering*, from Nanyang Technological University Singapore (www.ntu.edu.sg). I completed my high schooling from the National Public School in Bangalore in India in Jul-1999.